\documentclass[twocolumn,groupedaddress,prb,aps,showpacs,10pt]{revtex4-1}

\usepackage{amsmath,mathrsfs,amsbsy,color,graphicx,bm,amsthm,dsfont,amsfonts,afterpage,chngcntr,soul}
\usepackage[english]{babel}
\usepackage{amssymb,lipsum,attachfile}
\usepackage{amsmath,epsfig,epstopdf}
\usepackage{bibunits}
\usepackage{setspace}

\newcommand{\be}{\begin{equation}}
\newcommand{\ee}{\end{equation}}
\newcommand{\bea}{\begin{eqnarray}}
\newcommand{\eea}{\end{eqnarray}}
\newcommand{\ket}{\rangle}

\newcommand{\m}{\mathcal}
\newcommand{\X}{\sigma^x}
\newcommand{\Y}{\sigma^y}
\newcommand{\Z}{\sigma^z}

\begin{document}
\newtheorem{theorem}{Theorem}
\newtheorem{proposition}[theorem]{Proposition}
\newtheorem{corollary}[theorem]{Corollary}
\newtheorem{open problem}[theorem]{Open Problem}
\newtheorem{Definition}{Definition}
\newtheorem{remark}{Remark}
\newtheorem{example}{Example}

\title{Quantum Computing with sine-Gordon Qubits}
\author{Dong-Sheng Wang}
\email{wdscultan@gmail.com}
\affiliation{Institute for Quantum Computing and Department of Physics and Astronomy, \\University of Waterloo, Waterloo, Canada}

\begin{abstract}
A universal quantum computing scheme,
with a universal set of logical gates, is proposed based on networks of 1D quantum systems.
The encoding of information is in terms of universal features of gapped phases,
for which effective field theories such as sine-Gordon field theory can be employed to describe a qubit.
Primary logical gates are from twist, pump, glue, and shuffle operations
that can be realized in principle by tuning parameters of the systems.
Our scheme demonstrates the power of 1D quantum systems for robust quantum computing.
\end{abstract}

\date{\today}
\maketitle

\begin{spacing}{1.0}

Finding qubits with robust properties is crucial to build a quantum computer.
Robust quantum computing demands of good error-correction codes~\cite{Ter15},
which could accompany a large overhead,
or a self-correcting quantum memory~\cite{BLP+16},
which, as the analog of classical bits,
with the 2D Ising models as a physical cornerstone for classical computers,
is still missing.
Topological quantum computing (TQC)~\cite{Kit03},
with qubits usually carried by edge modes or anyonic excitations~\cite{MCA+14,AHM+16,BLK+17},
has been one of the most promising schemes for quantum computing.

Topological orders~\cite{Wen04},
which underlie the physics of TQC,
have seen significant progress in recent years.
This motivates the search for new schemes of TQC with distinct features.
With symmetry-protected topological (SPT) order~\cite{CGW11,SPC11},
1D quantum systems have been recognized as
promising candidates of quantum computer hardware,
both bosonic and fermionic~\cite{MCA+14,AHM+16,PCG+12,RAE+18,WAR18,Wang18}.
In this work, we study universal quantum computing
with qubits encoded in the bulk states of 1D gapped quantum spin systems.
With valence-bond solids and bosonization theory~\cite{AKLT87,Aff90,Gia04,GNT04},
robust code properties have been demonstrated recently~\cite{WAR18,Wang18}.
However, it is not known whether 1D gapped quantum spin systems,
and in general systems that can be well described by
sine-Gordon field theory and its equivalence~\cite{GNT04}
can support \emph{universal} quantum computing.
We achieve this by the design of a scheme with
a universal set of gate operations on such sine-Gordon qubits.

\begin{figure}[b!]
  \centering
  \includegraphics[width=.45\textwidth]{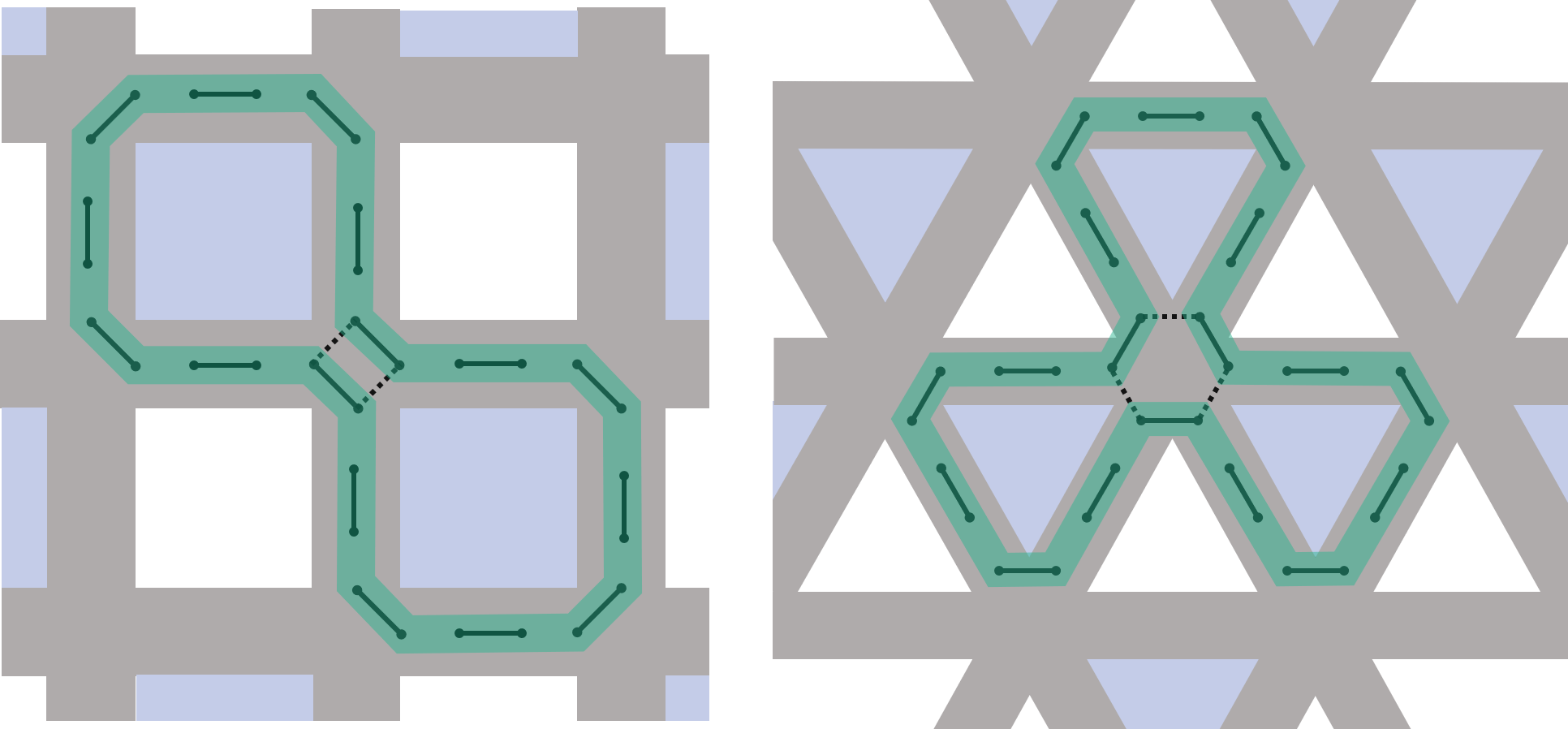}
  \caption{Networks of sine-Gordon qubits.
  1D systems (gray lines) are connected in networks forming
  square lattice (left) or triangular lattice (right).
  Each qubit is formed by edges of shaded plaquette.
  Unshaded plaquettes do not encode qubits.
  Single qubit gates are from operations on edges and plaquette,
  while entangling gates also involve glue operations on vertices.
  The (green) loops represent glued configurations of multiple qubits
  when the quantum switch at a shared corner converts the singlets from the initial ones (dashed) to the final ones (solid).
   }
   \label{fig:sGQC}
\end{figure}

In our scheme, a qubit is encoded in the universal property of the bulk states of gapped phases,
with field variables described by the sine-Gordon theory.
Logical gates follow from the so-called vertex algebra of field observable,
which have a certain topological robustness.
In particular, the logical phase flip $Z_L$ is a flux insertion
that can be realized by external global fields,
bit flip $X_L$ is a pump process of excitations in a cycle on the system.
Hadamard gate $H_L$, which exchanges $Z_L$ and $X_L$,
can be induced by the unitary shuffle process from one gapped phase to its dual,
or alternatively, by the quantum teleportation method~\cite{ZLC00}.
Entangling gates are from glue operations of qubit states
that can be realized by tuning of interaction parameters of a model.
The scheme is scalable forming various networks, as shown in Fig.~\ref{fig:sGQC}.
With the implementation in a spin-ladder system,
our scheme generalizes the classical magnetic logic,
and also extends the interplay between spintronics and quantum computing~\cite{LD98,MLL03,BMT+18}.
With bosonization,
our method shall be adapted to other systems~\cite{Kar86,PU95,OT01,WLF03,GV17,PPR+17},
including Josephson junctions arrays and lattice boson ladders,
which can be treated as quantum simulators,
hence serves as potential testbed
with advanced control technique for the scheme we propose.

We start from basic sine-Gordon field theory
and explain how it can describe a qubit,
and then study examples in 1D quantum spin system.
A simple sine-Gordon Hamiltonian takes the form
\be  H= H_0 + V(\phi),\ee
for a free Gaussian part $H_0$~\cite{Gia04,GNT04} and
a sine-Gordon nonlinear term $V(\phi)= g\int dx \cos \beta \phi$
with real parameters $g$ and $\beta$.
Here $x$ is the spatial direction along the system.
It describes the dynamics of conjugate bosonic field operators $\phi$ and $\theta$
(with the hat symbol omitted) such that
\be [\phi(x),\theta(y)]=i\Theta(x-y)\ee
for Heaviside step function $\Theta$.
The free part is massless while a mass can be induced if the nonlinear term is relevant
under renormalization flow~\cite{Gia04,GNT04}.

It is appropriate to understand the essence of the model as a harmonic oscillator with nonlinearity,
while the fields $\phi$ and $\theta$ are compactified (i.e., periodic).
As a result, $\phi$ and $\theta$,
or precisely, their values on the code space,
can be treated as the angular coordinates for an encoded qubit,
whose state can be expressed as
\be \rho\propto \mathds{1}+\vec{n}\cdot \vec{\sigma}, \ee
with Bloch vector
$\vec{n}=(\sin \phi \cos\theta, \sin\phi \sin\theta, \cos\phi )|\vec{n}|$,
and Pauli vector
$\vec{\sigma}=(\X,\Y,\Z)$.
The norm $|\vec{n}| \in (0,1]$ and stays the same under unitary transformation.
A state can be determined by the measurement of Pauli observables in $\vec{\sigma}$,
each of which is both unitary and self-adjoint.
Furthermore, in the framework of field theory physical observable are the so-called vertex operators~\cite{Gia04,GNT04}
$e^{i a \phi(x)}$ and $e^{i b \theta(y)}$,
$a,b\in \mathbb{R}$, and they satisfy
\be e^{i a \phi(x)} e^{i b \theta(y)}=e^{-iab \Theta(x-y)} e^{i b \theta(y)} e^{i a \phi(x)}, \label{eq:weyl}\ee
which is a Weyl algebra serving as
the physical foundation for logical operators on the qubit.

To proceed further, we recall the encoding of a qubit via dimerized states of spin-$\frac{1}{2}$
Heisenberg model with staggered or 2nd nearest-neighbor (NN) exchange interactions
with periodic boundary condition (PBC)~\cite{Wang18}.
A dimer is also known as a singlet,
and a dimerized state is a product of NN singlets.
The dimerization is due to breaking of lattice translation by odd number of sites,
denoted by $T$.
The ground state degeneracy (GSD) is two,
and a qubit can be encoded.
The two primary logical operators are
\be X_L=\left(\begin{matrix}
  0 & 1\\ 1 & 0 \end{matrix}\right),\;
   Z_L=\left(\begin{matrix}
  1 & 0\\ 0 & -1 \end{matrix}\right), \ee
known as bit-flip and phase-flip gates, respectively.
It is easy to see that the logical bit-flip
$X_L$ is $T$, as the generator of the broken translation symmetry.
Physically, $X_L$ can also be viewed as the pump of a spinon excitation along the system,
which could be interpreted as a Wilson loop of spinon,
and implementable by an adiabatic pumping cycle~\cite{Shi05}.
The logical phase-flip
$Z_L$ is the so-called twist operator~\cite{LSM61}
\be \label{eq:twist} F=\otimes_{n=1}^L e^{i\frac{2\pi}{L}nS^z_n},\ee
which extracts the SPT order of the ground states.
The $F$ operator is equivalent to a vertex operator of $\phi$~\cite{NV02,NT02,Wang18},
and can also be viewed as the insertion of a flux through the hole encircled by the system.
The algebra of $X_L$ and $Z_L$ is due to (\ref{eq:weyl}),
with $T$ equivalent to a vertex operator of $\theta$.

For universal quantum computing,
a universal set of quantum gates are required.
For sine-Gordon qubits,
the apparent difficulty is that there exists a discord between
$X_L$ and $Z_L$, namely,
they are from different mechanism of symmetries,
as discussed above.
This means there is no easy way to realize the logical Hadamard operator
\be H_L=\frac{1}{\sqrt{2}}\left(\begin{matrix}
  1 & 1\\ 1 & -1 \end{matrix}\right), \ee
which switches $X_L$ and $Z_L$ and generates superposition.
We find that this difficulty can be resolved by
employing more sophisticated spin systems and the mechanism of \emph{duality}.

To put duality in the setting of quantum computing,
first consider the encoding of a classical bit by the 2D Ising model.
Below the critical temperature $T_c$,
there are two subspaces, denoted $\m{C}_0$, $\m{C}_1$,
due to the breaking of a global $Z_2$ symmetry.
The bit 0 (1) is encoded as $\m{C}_0$ ($\m{C}_1$) with total magnetization $M$ up (down),
and the logical bit flip $X_L$ is from the broken $Z_2$ symmetry.
This code is known as a repetition code with codewords determined by the majority-vote rule.
Furthermore, the code is a subsystem code~\cite{NC00} in the following sense.
The whole space $\mathcal{H}$ can be decomposed as
\be \mathcal{H} \cong \m{C}_0 \oplus \m{C}_1 \cong \mathbb{C}^2 \otimes \mathcal{G}.
\label{eq:isings}\ee
The code space is $\mathbb{C}^2$ and the rest is a so-called gauge space $\mathcal{G}$.
As the dimension of $\mathcal{H}$ is even,
states in $\m{C}_0$ and $\m{C}_1$ are one-to-one correspondent,
leading to the code space $\mathbb{C}^2$ encoding the sign of $M$.
Local thermal noises that do not flip the sign of $M$ are described by $\mathcal{G}$.

Now, if we treat it as a qubit and consider the superposition
$|\pm\ket=\frac{1}{\sqrt{2}}(|0\ket\pm|1\ket)$,
$M$ is basically zero.
This indicates that $|\pm\ket$ have different order from $|0,1\ket$.
To be more concrete, consider the quantum version of the 2D Ising model
\be H=-\sum_n \X_n - \lambda \sum_n \Z_n \Z_{n+1}. \ee
The critical point is $\lambda_c=1$,
and large (small) $\lambda$ corresponds to the low (high) temperature phase.
A notable feature is the duality~\cite{FS78,Kog79,CON11} defined by
$\tilde{\sigma}^x_n=\Z_{n}\Z_{n+1}$, $\tilde{\sigma}^z_n=\prod_{m<n}\X_{m}$ on the dual lattice,
from which
$H=- \lambda \sum_n \tilde{\sigma}^x_n -  \sum_n \tilde{\sigma}^z_n \tilde{\sigma}^z_{n+1}.$
The high-temperature phase has zero order $\sum_n \Z_n$ while
a nonzero `disorder' $\sum_n \tilde{\sigma}^z_n$,
and the opposite for the low-temperature phase~\cite{Fra17}.

In fact, the duality serves as
the logical Hadamard gate
$H_L$ that switches between the order and disorder.
Namely, $H_L: X_L \leftrightarrow Z_L$ for $Z_L=\Z_n$ and $X_L=\otimes_n \X_n$.
This means the space of the high-temperature phase also divides into two parts,
$\m{C}_+$ and $\m{C}_-$ for the even and odd parity of $X_L$.
Note the parity can be viewed as the number of local states $|-\ket$
in a configuration as a product of local $|+\ket$ and $|-\ket$ states,
assuming the system size is odd without loss of generality.
Suppose that there is no noise term $\Z_n$ on an odd number of sites
and $\lambda$ can be tuned properly,
then the exchange of the two phases implements $H_L$.
However, a phase flip $\Z_n$ on a local site $n$ can be easily induced in practice,
which is the key reason for it being only a good classical bit.

Now we study models that can be described by the sine-Gordon field theory
and provide qubit with robust logical $X_L$ and $Z_L$, and $H_L$ from duality.
This would surpass the encodings via Ising model or spin chains discussed above,
and integrate SPT order with duality for better encoding of qubits.
We find the two-leg spin-$\frac{1}{2}$ ladder is a system with duality property,
as a natural extension of the single leg case.
With the new fields
\be \phi_\pm=\frac{1}{\sqrt{2}}(\phi_1\pm\phi_2),\;
\theta_\pm=\frac{1}{\sqrt{2}}(\theta_1\pm\theta_2), \ee
and $1,2$ labeling the two legs,
we employ the Hamiltonian
\be H= H_0 + V(\phi_+) + V(\phi_-) + V(\theta_-) \label{eq:sgHa}\ee
for the free part of the ladder $H_0$,
the symmetric part $V(\phi_+)=g_1 \int dx \cos 2\sqrt{\pi} \phi_+$,
and antisymmetric parts $V(\phi_-)=g_2 \int dx \cos 2\sqrt{\pi} \phi_-$,
$V(\theta_-)=g_3 \int dx \cos 2\sqrt{\pi} \theta_-$.
Terms with $\theta_+$ are forbidden by the global symmetry.
Coupling constants $g_i$ are functions of original exchange strengths, and
the phase diagram is well established~\cite{SNT95,NT97,KFS+99,WLF03,SB04,VH06,KLS08}.
There are four gapped phases, as summarized in Table~\ref{tab:phase}.
Columnar (C) dimer phase: its ground states are dimerized with aligned dimers, and the GSD is two.
Staggered (S) dimer phase: its ground states are dimerized with staggered dimers, and the GSD is two.
Haldane (H) phase: the ground state is unique, and each two aligned spins form an effective triplet,
and there is a singlet between each two triplets.
Rung-singlet (R) phase: the ground state is unique, and each two aligned spins form a singlet.

\begin{table}
  \centering
  \begin{tabular}{c|c}
    \hline
    $(\phi_+,\phi_-)$ & $(\phi_+,\theta_-)$ \\ \hline
    $|0,1\ket$ \hspace{.4cm} C & $|\pm\ket$ \hspace{.4cm} R  \\ \hline
    \hspace{1.2cm} S & \hspace{1.1cm} H  \\
    \hline
  \end{tabular}
  \caption{Encoding of qubit in the phases of the two-leg spin-$\frac{1}{2}$ ladder.
  Logical states $|0,1\ket$ are encoded in the columnar (C) dimer phase,
  and $|\pm\ket$ in the Rung-singlet (R) phase.
  The staggered (S) dimer phase and Haldane (H) phase form another equivalent code space.
  The C and S phases are specified by fields $\phi_+$ and $\phi_-$,
  and R and H phases by fields $\phi_+$ and $\theta_-$.
   }\label{tab:phase}
\end{table}

For the phases C and S, the two spin chains are decoupled,
while for phases H and R, the two spin chains are coupled, and in fact, entangled.
The phases C and S each have definite values of $\phi_+$ and $\phi_-$,
and phases H and R each have definite values of $\phi_+$ and $\theta_-$.
Viewed as a continuous-variable system~\cite{EPR35}
analog with harmonic oscillator or photonic states,
low-energy states of phases H and R each are entangled with respect to $\phi_{1,2}$ and $\theta_{1,2}$.
This motivates the encoding of a qubit as follows.
The state space of phase C is divided into two parts $\m{C}_0$ and $\m{C}_1$ due to dimerization,
and the codeword $|0\ket:=\m{C}_0$ ($|1\ket:=\m{C}_1$) with positive (negative) values of $\cos \sqrt{2\pi}\phi_-$.
The codewords $|+\ket:=\m{C}_+$ ($|-\ket:=\m{C}_-$)
with positive (negative) values of $\cos 2\sqrt{\pi}\theta_-$ of phase R.

Note that a field theory describes low-energy,
including low-lying excitations, universal features of the spin system.
To characterize the logic of quantum computing,
it is a great advantage of using field theories since
quantum information is encoded in the universal features of phases.
The total space of a spin system is decomposed as
\be \mathcal{H} \cong \m{C} \oplus \m{C}^{\perp},\;
 \m{C} \cong \m{C}_0 \oplus \m{C}_1 \cong \mathbb{C}^2 \otimes \mathcal{G}.
\label{eq:sgs}\ee
The space $\m{C}^{\perp}$ represents the high-energy part that cannot be well described by
sine-Gordon field theory,
which plays a trivial role for the Ising case~(\ref{eq:isings}).
The code space $\mathbb{C}^2$ is due to dimerization,
and it also has SPT order that is absent for the Ising model.
The gauge space $\mathcal{G}$ can be interpreted as the part
for soliton excitations (and their bound states) that will not make a logical error.

The encoding via the ladder system
can be viewed as a repetition code concatenated with an underlying code by a single chain.
In the language of stabilizer code~\cite{NC00},
which describes codes that are stabilized by a set of commuting operators,
the codewords $|0\ket$ and $|1\ket$ are stabilized by $Z_1Z_2$ while $Z_L$ is $Z_1$ or $Z_2$,
and $|\pm\ket$ are stabilized by $Z_1Z_2$ and $X_L=X_1X_2$.
In the spin language, and let $\phi$ be $\phi_1$ or $\phi_2$, the logical operators are
\be X_L:=e^{i2\sqrt{\pi}\theta_-},\; Z_L:= e^{i\sqrt{2\pi}\phi}.\ee
The logical bit flip $X_L$ is a pump operation of spinons,
and the logical phase flip $Z_L$ is a twist or flux operation
and can be realized by inserting electric fields~\cite{Wang18}.

The logical Hadamard gate $H_L$ is played by the duality mapping between $X_L$ and $Z_L$,
which can be realized by slowly tuning parameters, e.g., $g_2$ and $g_3$,
in the Hamiltonian~(\ref{eq:sgHa}) as a unitary process
shuffling between phases C and R.
For large but finite system sizes,
the shuffle operation can be engineered to be unitary in principle,
and it will map between the corresponding eigenspaces of $X_L$ and $Z_L$.
In the thermodynamic limit,
the gap-closing during the shuffle labels
the change of order parameters (see Table~\ref{tab:phase}), i.e.,
the 2nd order phase transition between phases C and R.
The phase transition could jeopardize the exact reversibility of the shuffle operation in practice.
However, as long as thermal noises do not lead to logical errors $X_L$ or $Z_L$,
the shuffle realizes the unitary gate $H_L$ on the logical level.
Also as the encoding is via low-lying subspaces instead of merely ground states,
the system does not have to be maintained on ground states.

Another common method to realize gates is by gate teleportation~\cite{ZLC00}.
For the Hadamard gate $H_L$, it requires an entangling gate and projective measurement.
With the CZ gate, a scheme for which is explained below,
the $H_L$ can be realized as
\be \langle m|_\textsc{s} H_\textsc{s} CZ |\psi\rangle_\textsc{s} |+\rangle_\textsc{a}= X^m H |\psi\rangle_\textsc{s}, \; m=0,1,\ee
given an arbitrary qubit state $|\psi\rangle_\textsc{s}$,
and a qubit ancilla prepared on state $|+\rangle_\textsc{a}$,
and the projective measurement on X basis $\langle m|_\textsc{s} H_\textsc{s}$ on the qubit.
The byproduct $X^m$ can be corrected given the measurement outcome $m$,
and the output is the state $H |\psi\rangle_\textsc{s}$.

Our encoding is similar but greatly generalizes that for the Ising model.
The phases employed here not only support symmetry breaking,
but also have symmetry-protected topological order.
The global logical operators $X_L$ and $Z_L$
detect the proper topological order parameters of these phases,
and hence, they are not easy to be mimicked by the noisy environment.

In addition, phases S and H form another `copy' of code space,
which differs from the original code space of phases C and R by the value of $e^{i2\sqrt{\pi}\phi_+}$,
which has definite values for all the phases.
The observable $e^{i2\sqrt{\pi}\phi_+}$ flips its sign
when phases C and S (or R and H) are exchanged,
by, e.g., $T$ on one of the two spin chains,
which is a 1st order phase transition and can be detected.
In practice, to locate a disturbed chain,
a slight asymmetry can be introduced for the two legs of the ladder
so that the two legs can be distinguished.
The correction is then the operation $T$ itself, or the pump of spinon along the disturbed chain.
The measurement of $e^{i2\sqrt{\pi}\phi_+}$ also
benefits initialization and the entangling gates.
Preparation of a logical state can be done by cooling
and energy splitting from staggered interaction, for instance.
To identify a logical state,
hermitian inter-chain or intra-chain dimer order parameters can be measured.

For universal quantum computing,
entangling gates are required.
Next we propose a method to realize the well-known CZ gate and CCZ gate.
The CZ (CCZ) gate generates a minus sign when the two (three) qubits are on logical state $|1\ket$.
For convenience, we denote $CZ\equiv \Lambda_2$, $CCZ\equiv \Lambda_3$,
and it will be clear that our method can also be employed to realize $\Lambda_n$ for $n>3$,
which, however, are great challenges for control technique.

In the setting of TQC,
our method to realize entangling gates is by the change of topology.
This is to glue (or merge) loops of states for 1D systems with PBC.
As the states of a qubit can be properly viewed as
loops of singlets except a few excitations,
states from different qubits can be glued together,
which is a topological quantum operation and enables entangling gates.
Therefore, qubits can be arranged on 2D lattices with `point contact' between all NN pairs,
see Fig.~\ref{fig:sGQC} for the square lattice and triangular lattice.
A controllable interaction at a corner, as a quantum `switch',
glues qubits together conditioned on special states of them.
An entangling gate $\Lambda_n$ is realized by the sequence of
glue, a global twist, and then deglue.

\begin{figure}
  \centering
  \includegraphics[width=.45\textwidth]{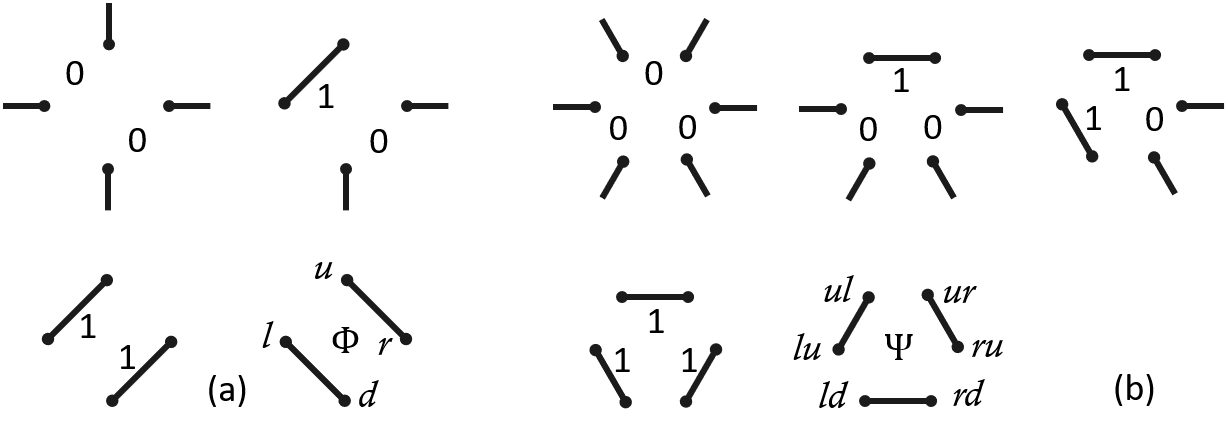}
  \caption{The singlet configurations at the corner of the 2D square lattice (a) and triangular lattice (b).
  A spin-$\frac{1}{2}$ is shown as a dot, and a singlet as a bar.
  The 0 (1) labels logical state $|0\ket$ ($|1\ket$) of a qubit.
  The state $|01\ket$ is a rotated version of $|10\ket$ (a),
  states $|001\ket$, $|010\ket$ are rotated versions of $|100\ket$,
  and $|011\ket$, $|011\ket$ are rotated versions of $|110\ket$ (b).
  The glued states are $|\Phi\ket$ (a) and $|\Psi\ket$ (b).
  The corner sites are labeled as u(p), d(own), l(eft), r(ight) (a),
  ul, ur, lu, ld, ru, and rd (b).}
  \label{fig:corner}
\end{figure}

We now show the details for the spin ladder system.
The model~(\ref{eq:sgHa}) can be realized by two-leg spin-$\frac{1}{2}$ ladder
with spin exchange interaction $\vec{S}_{i}\cdot \vec{S}_{j}$
for NN (and possible 2nd NN) sites on each chain,
each pair of sites on the rungs and along plaquette diagonals~\cite{SNT95,NT97,KFS+99,WLF03,SB04,VH06,KLS08}.
First, to illustrate the basic mechanism,
consider the case when a single chain is used for a qubit.
At a corner of a square lattice,
there are four spins, for which the five proper singlet configurations are shown in Fig.~\ref{fig:corner}(a),
and the two qubits sit at the northwest and southeast plaquette.
They represent logical states
$|00\ket$, $|01\ket$, $|10\ket$, and $|11\ket$ of the two qubits,
and the last one is a glued state, denoted as $|\Phi\ket$.
Denote the sites of the four spins as u(p), d(own), l(eft), r(ight),
an exchange interaction $\vec{S}_u\cdot \vec{S}_l$ as $h_{ul}$,
we employ an antiferromagnetic interaction
\be H=J( h_{ul} + h_{dr})+ J_g( h_{ur} + h_{dl}). \ee
The ratio $J_g/J$ shall be adiabatically tuned to a big value such that
the state $|11\ket \mapsto |\Phi\ket$
while others stay the same.
Now the glued system can be viewed as a whole system,
with states $|00\ket$ and $|\Phi\ket$ serving as the two new degenerate dimerized logical states.
A global twist similar with (\ref{eq:twist}) on the two qubits will enable
\be  |00\ket\mapsto |00\ket,  |\Phi\ket \mapsto -|\Phi\ket.\ee
The states $|01\ket$ and $|10\ket$
each has two domains and the two spinons at the domain walls forming a singlet at the corner,
hence are low-lying excited states above $|00\ket$ and $|\Phi\ket$.
The global twist will break the corner singlets leading to modified states $|01'\ket$ and $|10'\ket$,
which will acquire the same dynamical phase $e^{i\delta}$,
$\delta=Et$ for their energy $E$ and free evolution time $t$.
After the twist, the deglue operation will drive back to the code space, namely,
$|\Phi\ket \mapsto |11\ket$, $|01'\ket \mapsto |01\ket$,
$|10'\ket \mapsto |10\ket$, and $|00\ket$ stays the same.
If $t$ is short enough such that $e^{i\delta}\approx 1$,
the sequence of glue-twist-deglue (GTG) enables the gate $\Lambda_2$.
Further, it is not hard to see for the spin-$\frac{1}{2}$ ladder,
with the two legs arranged along the third dimension, i.e., vertically,
and eight spins at a corner,
the same mechanism works leading to the gate $\Lambda_2$.

The GTG scheme can be applied to the triangular lattice to implement the gate $\Lambda_3$,
where there are twelve spins at a corner arranged as two diamonds overlapped vertically.
For each six spins, labelled as ul, ur, lu, ld, ru, rd,
we employ the interaction
\be \label{eq:glue}
H=J( h_{ul,ur} + h_{lu,ld} +h_{ru,rd})+J_g( h_{ul,lu} + h_{ur,ru} + h_{ld,rd}).
\ee
Now there are nine proper singlet configurations, shown in Fig.~\ref{fig:corner}(b).
The glue interaction~(\ref{eq:glue}) will map between states $|111\ket$ and $|\Psi\ket$
by tuning the value $J_g/J$.
The $\pi$ phase shift on $|\Psi\ket$ is induced by the global twist on the three qubits.
During the GTG operation, state $|000\ket$ stays the same,
while $|100\ket$, $|010\ket$, and $|001\ket$ obtain the same phase $e^{i\delta_1}$,
$|110\ket$, $|101\ket$, and $|011\ket$ obtain the same phase $e^{i\delta_2}$,
both of which can be made trivial by reducing the time of free evolution.
Overall, by the GTG operation the gate $\Lambda_3$ can be realized.
It is also clear that the gate $\Lambda_2$ can be realized on this lattice.
As the result, this system supports the universal gate set $\{H_L, \Lambda_3\}$~\cite{Shi02},
hence can be used for universal quantum computation.

Our study serves as a constructive proof of the universality (and scalability) of sine-Gordon qubits,
demonstrating the power of (quasi-)1D quantum systems
for quantum computation.
To realize this, there are great practical challenges.
As a qubit is encoded in the two-leg ladder system,
the bit-flip and phase-flip operations may require controllability of any single leg.
Although bit-flip and phase-flip gates can be realized by global operations,
the current proposal of Hadamard gate and entangling gates rely on
tunability of interaction terms.
The shuffle between phases for the Hadamard gate may only be realized approximately due to noises.
The teleportation scheme avoids this subtilty,
yet it requires the entanglement with an ancilla and projective measurement.
The entangling gates $\Lambda_n$ require precise timing and local addressability (at the corner).
A global scheme to realize entangling gates would be appealing,
which remains as an interesting open question.

To summarize,
a scalable topological quantum computing scheme based on spin ladder,
and sine-Gordon qubits in general, is proposed.
The computation shall be robust against a certain perturbation of control parameters
as qubits are encoded into phases instead of states.
The lifetime of a qubit, although topological, shall be affected by various control process.
Our scheme reveals a novel relation between quantum computing and phase transition.
Our method can also be extended to multi-leg or high-spin ladders
and other relevant systems.

\emph{Acknowledgement.--}This work has been funded by NSERC.

\end{spacing}

\bibliography{ext}{}

\end{document}